\documentclass[12pt]{iopart}
\usepackage{iopams}
\usepackage{graphicx}

\newcommand{\CD}{{\cal D}}
\newcommand{\CR}{{\cal R}}
\newcommand{\CQ}{{\cal Q}}

\newcommand{\I}{{I}}
\newcommand{\II}{{II}}

\newcommand{\average}[1]{\left\langle #1 \right\rangle_\CD}
\newcommand{\laverage}[1]{\left\langle #1 \right\rangle_{\CD_{\rm \bf i}}}

\newcommand{\naverage}[1]{\left\langle #1 \right\rangle_{\CD_{\rm \bf 0}}}
\newcommand{\initial}[1]{{#1_{\rm \bf i}}}
\newcommand{\now}[1]{{#1_{\rm \bf 0}}}
\newcommand{\be}{\begin{equation}}
\newcommand{\ee}{\end{equation}}
\newcommand{\bea}{\begin{eqnarray}}
\newcommand{\eea}{\end{eqnarray}}
\newcommand{\bean}{\begin{eqnarray*}}
\newcommand{\eean}{\end{eqnarray*}}
\begin{document}

\title[Testing backreaction effects with observations]{Testing backreaction effects with observations}
\author{Julien Larena$^{1,2}$, Jean--Michel Alimi$^{2}$, Thomas Buchert$^{3,2,4}$, Martin Kunz$^{4,5}$, and Pier-Stefano Corasaniti$^{2}$}
\address{$^1$Cosmology and Gravity Group, Department of Mathematics and Applied Mathematics, University of Cape Town; Rondebosch 7700 Cape Town, South Africa} 
\address{$^2$CNRS, Laboratoire Univers et Th\'eories (LUTH), CNRS UMR 8102, Observatoire de Paris, Universit\'e Paris Diderot ; 5 Place Jules Janssen, 92190 Meudon, France}
\address{$^3$Universit\'e Lyon 1, Centre de Recherche Astrophysique de Lyon (CRAL), CNRS UMR 5574, 
9 avenue Charles Andr\'e, F--69230 Saint--Genis--Laval, France}
\address{$^4$D\'epartement de Physique Th\'eorique, Universit\'e
de Gen\`eve, 24 quai Ernest Ansermet, CH--1211 Gen\`eve 4,
Switzerland}
\address{$^5$Astronomy Centre, University of Sussex, Falmer, Brighton BN1 9QH, UK}
\smallskip
\address{{\it Emails:} julien.larena@obspm.fr$\,,\,\,$jean-michel.alimi@obspm.fr$\,,\,\,$buchert@obs.univ-lyon1.fr$\,,\,\,$m.kunz@sussex.ac.uk$\,,\,\,$pier-stefano.corasaniti@obspm.fr}
\begin{abstract}
In order to quantitatively test the ability of averaged inhomogeneous cosmologies to correctly describe observations of the large scale properties of the Universe, we introduce a smoothed template metric corresponding to a constant spatial curvature model at any time, but with an evolving curvature parameter. This metric is used to compute quantities along an approximate effective lightcone of the averaged model of the Universe. 
As opposed to the standard Friedmann model, we parameterize this template metric by exact 
scaling properties of an averaged inhomogeneous cosmology, and we also motivate 
this form of the metric by results on a geometrical smoothing of inhomogeneous cosmological
hypersurfaces. The purpose of the paper is not to demonstrate that the backreaction effect is actually responsible for the Dark Energy phenomenon by explicitly calculating the effect from a local model of the geometry and the distribution of matter, but rather to propose a way to deal with observations in the backreaction context, and to understand what kind of generic properties have to hold in order for a backreaction model to explain the observed features of the Universe on large scales. 
We test our hypothesis for the template metric against supernova data and the position of the CMB peaks, and infer the goodness--of--fit and parameter uncertainties. We find that averaged inhomogeneous models can reproduce the observations without requiring an additional Dark Energy component (though a volume acceleration is still needed), and that current data do not disfavour our main assumption on the effective lightcone structure. We also show that the experimental uncertainties on the angular diameter distance and the Hubble parameter from Baryon Acoustic Oscillations measurements -- forseen in future surveys like the proposed EUCLID satellite project -- are sufficiently small to distinguish between a FLRW template geometry 
and the template geometry with consistently evolving curvature.

\end{abstract}
%

\section{Introduction}
\label{sec:intro}

On the very large scales the Universe appears to be close to
a homogeneous and isotropic state. This is usually modeled by a
locally isotropic and hence homogeneous solution of Einstein's
equations, namely the standard Friedmann--Lema\^\i tre--Robertson--Walker (FLRW) metric.
The observations of tiny temperature fluctuations of the Cosmic
Microwave Background (CMB) radiation suggest that in the Early
Universe deviations from this `background cosmology' were very small, 
thus motivating the use of linear perturbation theory about the FLRW solution.
The experimental confirmation of the predicted baryon acoustic oscillations in
the CMB power spectrum \cite{debernardis,spergeletal}, as well as the distribution of galaxies and clusters on the
large scales \cite{lahav} provide a certain body of evidence in favor of the
`concordance model' which results from this perturbative
approach (see, however, \cite{marraPN}, \cite{huntsarkar}). This standard scenario relies on the assumption that the FLRW
cosmology correctly describes the `background cosmology', i.e. the
averaged inhomogeneous Universe,  at all times. Even though this hypothesis 
may be valid in the Early Universe, does it continue to hold even when the Universe
becomes more and more structured at late times? 

Answering such a question has become even more important in the light of the still
unexplained Dark Energy phenomenon in the context of the FLRW paradigm.
The luminosity distance measurements to type Ia supernova (SN-Ia) standard candles, 
when analyzed within the framework of the FLRW Universe,
provide strong evidence for a missing component characterized by a negative pressure
which, by inducing an accelerated phase of expansion, would be responsible for
the observed dimming of far distant SN Ia (\cite{Perlmutter,Riess}). 
The simplest scenario to account for these observations 
is a positive cosmological constant in Einstein's equations. This is often assumed to
describe the energy contribution
of quantum vacuum fluctuations. Nevertheless, because of the huge discrepancy between the particle
physics expected value and the observed one, several alternative scenarios have been investigated.
For instance phenomenological models such as a late time slow rolling scalar field 
(see reviews \cite{copeland,pilar}) or the Chaplygin gas \cite{chaplygin} have been proposed
to describe this Dark Energy component.
Alternatively, there have been several proposals to account for these effects through modification of
the laws of gravitation (e.g. braneworlds \cite{maartens:brane}, scalar--tensor gravity \cite{esposito}, higher--order gravitational theories \cite{riccilagrangians,capo}, AWE \cite{AWE1,AWE2}).  

Recently, a third alternative has been considered
\cite{rasanen:darkenergy,kolbetal} that aims at explaining Dark Energy as
an effect caused by inhomogeneities. 
However, most of the approaches which include the effect of
inhomogeneities still rely on the postulate that the FLRW solution reliably describes the effective (average) evolution of 
an inhomogeneous cosmology. For instance this is the case of
inhomogeneous universe models in which distances are computed using perturbation
theory about a FLRW background \cite{ruth:luminosity,vanderveldetal}. 
Other efforts abandon the FLRW model and instead restrict inhomogeneities by strong symmetries,
employing exact solutions to Einstein's field
equations like the Lema\^\i tre--Tolman--Bondi (LTB) metric 
\cite{celerier1, rasanen:LTB, nambu, LTBgron, LTBluminosity5, LTBluminosity1, LTBluminosity2, LTBluminosity4, alnes2, alnes3, singh1,Bolej,sussman2}. 

In this work we shall test a different approach and exploit the key--insight that
the (large--scale) kinematics of a homogeneous--isotropic {\em state} does not necessarily follow the kinematics of 
a homogeneous--isotropic {\em solution}, especially at late epochs
characterized by the presence of large matter inhomogeneities. 
Indeed, the analysis of backreaction effects due to inhomogeneities suggests that there
is a wider class of  (large--scale) homogeneous--(almost--)isotropic cosmological models, while smaller scales 
feature strong inhomogeneities and anisotropies that both are known to exist. 
In such a case it is natural to ask whether the emergence of
Dark Energy can result from the breakdown of the underlying assumption associated with the FLRW cosmology. 

If no assumption on the nature of the inhomogeneities is made, i.e. if we do not restrict them to be 
small deviations from a FLRW background or obeying strong symmetry restrictions,
 we can still look at effective (average) properties of Einstein's equations.
In the simplest case such a programme can be realized by foliating spacetime into flow--orthogonal hypersurfaces, 
restricting the matter model to `dust', and
spatially averaging the scalar parts of Einstein's equations with respect to a collection of free--falling observers (generalized fundamental observers).
Whereas such a formalism and the dynamical equations that govern the
behavior of the averaged inhomogeneous universe model are well--established \cite{buchert:review}, the explicit 
geometry, which lies at the basis of how we measure distances, is left unspecified.

Here we suggest, as a next step, to complement the general kinematical properties of an averaged universe model with an
explicit form of a {\em template metric} that retains the
main properties of the standard model of cosmology, such as its isotropy and homogeneity on large scales, 
but allows for structuring on small scales. The shift in emphasis is from postulating a {\em strong cosmological principle} 
that assumes local isotropy about every point and hence homogeneity on all scales, to a
{\em weak cosmological principle} that only assumes (quasi--) isotropy and homogeneity on the largest observable scales.
In this context, we retain by assumption the usual description of the Universe at early times, up to decoupling. However, at
late times we will have to modify this description.

To summarize let us list our main goals and assumptions. The purpose of this paper is to study the influence of a geometrical effect induced by the coupling between backreaction and averaged spatial curvature, on top of the
well--studied kinematical effect of backreaction on the evolution of the effective volume scale factor \cite{buchert:review}. To realize this study, we propose an ansatz for the effective metric of the large-scale homogeneous model that is motivated by previous results on the smoothing of Riemannian metrics by the Ricci flow. This is used to define an effective background on which the photons propagate; such a background can be considered as a first refinement of the usual FLRW background geometry. In order to get an insight into the effects associated with our prescription, we consider a specific example of backreaction, namely a power law of the effective scale factor. Here we want to stress the fact that our aim is not to show that backreaction effects can be fully responsible for the Dark Energy phenomenon. Instead, we address the converse problem: what is necessary and what kind of generic effects are expected for a backreaction model to be consistent with the cosmological observations? And more specifically, we are interested in understanding what specificities can allow to distinguish between FLRW and averaged models.

The central concept underlying our investigation is that, although the 3--Ricci curvature
distribution of an inhomogeneous cosmological slice can be smoothed {\em at any time}
into a constant curvature, the dynamical evolution of the averaged curvature can
differ from the evolution of a constant--curvature (homogeneous) model.
This deviation has recently been quantified in the framework of perturbation theory \cite{lischwarz,lischwarz1,lischwarz2} (see also \cite{Ian} for an estimation in the conformal Newtonian gauge), and since we have arguments why a non--perturbative treatment is necessary for the effects of interest, we shall consider the dominant perturbative mode within a general class of scaling solutions to a backreaction--driven cosmology.
Of course, the scaling solutions cannot be expected to fully represent the realistic backreaction effect throughout the whole history of the Universe, but it is considered here for reasons of clarity and simplicity to illustrate the kind of effects expected from the non-trivial geometry, in analogy to studies using parameterizations of the equation of state for Dark Energy.

The paper is organized as follows. We introduce the backreaction
context, the key--equations and free parameters of an averaged
cosmological model in Section~2. In Section~3 we develop the ansatz for the metric, which is inspired by the study of the Ricci flow deformation of three--dimensional Riemannian
initial data sets, and we use this effective metric to compute quantities along an approximate past lightcone, in particular the luminosity distances of cosmological objects. This template metric can be considered as an improvement over the FLRW ansatz that considers that photons follow the null geodesics of a locally homogeneous and isotropic metric.
In Section~4 we discuss the constraints on the model parameters as inferred from SN Ia data and the 
multipoles of the CMB peaks and dips. In particular we calculate the cosmic distances in the template metric and determine the 
`best-fit' models with the help of exact scaling solutions to the
backreaction problem, which include the leading 
perturbative mode. We conclude this section with a discussion on possible tests of our main assumptions. 
Finally, we summarize the results of the paper and present an outlook in Section~5. 

\section{Backreaction, effective equations and their parameters}

\subsection{The backreaction context}

Let us first recall some of the main points of 
kinematical backreaction as a candidate for
Dark Energy.  
{\it Kinematical backreaction} comprises the averaged effect of 
inhomogeneities in matter and geometry, which lead 
to deviations of the kinematics of the averaged model from that of the
standard FLRW model. The existence of such deviations challenges the basic 
conjecture of the standard model, namely that the Universe is described
by a homogeneous--isotropic solution {\it on average}. 
A more general approach should encode deviations from the standard model
through kinematical backreaction terms
(see, e.g. \cite{rasanen,buchert:review} for summaries of the basic framework set out in
\cite{buchert:grgdust,buchert:grgfluid,buchert:static} and discussions
therein). The simplest account for this effect has
recently been summarized in terms of a {\it morphon field}, namely
a scalar field that provides an effective description of the
  backreaction effects within a standard Friedmannian framework \cite{morphon}.
More specifically, the price to pay of using Friedmann's equations to
describe the kinematics of a spatially averaged model, is the
presence of an extra component. 
In the FLRW context such a component would be attributed
to an exotic source (either a scalar field or a modification of
gravity) exhibiting (phantom--)quintessence like properties \cite{copeland}.

\subsection{Exact kinematics of the volume scale factor}
Let us consider a Universe filled with an irrotational perfect fluid of dust-matter with energy density $\varrho$. We foliate space--time using the ADM procedure \cite{ADM}, and restrict our analysis to the simple case where the lapse function is constant and the shift vector vanishes. This choice of local coordinates is equivalent to consider a family of observers comoving everywhere with the fluid. We are then left with a set of spatial hypersurfaces of 3--metric $g_{ij}(t,X^{i})$, parameterized by a universal time coordinate $t$, where $\left(X^{i}\right)_{i\in \{1,2,3\}}$ are Gaussian normal coordinates on the hypersurfaces associated with the special choice of comoving observers.
Following \cite{buchert:grgdust} we define an effective 
scale factor over a compact, rest mass--preserving domain $\CD$ that is contained within a spatial hypersurface orthogonal to the flow of the dust fluid,
\be
a_\CD(t) = \left(\frac{V_\CD (t)}{V_{\initial\CD}}\right)^{1/3} \;,
\ee
where $V_{\initial\CD} =\vert{\initial\CD}\vert$ denotes the initial value of the volume.
We also introduce an averaging operator that acts on any scalar function $\Psi$ as follows:
\begin{equation}
\label{eq:averagingop}
\average{\Psi}:=\frac{1}{V_{\CD}}\int_{\CD}\Psi Jd^{3}X \;\;,
\end{equation}
with the domains volume 
\begin{equation}
\label{eq:defvol}
V_{\CD}:=\int_{\CD}Jd^{3}X\;,
\end{equation}
and $J$ the square root of the 3--metric determinant in the spatial hypersurfaces.
Then, by averaging the Raychaudhuri equation, we obtain a second--order differential equation for the volume scale factor:
\begin{equation}
\label{averagedraychaudhuri}
3\frac{{\ddot a}_\CD}{a_\CD} + 4\pi G \average{\varrho} -\Lambda\;=\; {\CQ}_\CD\;,
\end{equation}
where the extra term ${\cal Q}_\CD$ is the kinematical backreaction term, 
 \begin{equation}
\label{Q} 
{\cal Q}_\CD \equiv 2 \average{\II} - \frac{2}{3}\average{\I}^2 =
\frac{2}{3}\average{\left(\theta - \average{\theta}\right)^2 } - 
2\average{\sigma^2}\;.
\end{equation}
$\I = \theta$  and $\II = 1/2[\,(\theta^2 - \Theta^i_{\;\,j}\Theta^j_{\;\,i}\,]$  
denote  the  principal scalar invariants  of the  expansion
tensor (here this tensor is symmetric and defined as minus the extrinsic curvature). 
We have written Eq.~(\ref{Q}) in terms of kinematical scalars through the decomposition  
$\Theta^i_{\;\,j} = \frac{1}{3}\delta^i_{\;j}\theta + \sigma^i_{\;\,j}$, 
where $\theta$ is the expansion rate, $\sigma_{ij}$ the shear tensor, 
$\sigma^2 \equiv 1/2 \sigma^{ij}\sigma_{ij}$ the rate of shear (summation over repeated indices
is understood).

Averaging the Hamiltonian constraint we obtain a further equation:
\begin{equation}
\label{averagedhamilton}
3\left( \frac{{\dot a}_\CD}{a_\CD}\right)^2 - 8\pi G \average{\varrho}-\Lambda \;=\; - \frac{\average{\CR}+{\CQ}_\CD }{2} \;,
\end{equation}
where $\CR$ is the 3--Ricci scalar of the spatial hypersurfaces. This equation is compatible with Eq.~(\ref{averagedraychaudhuri}), provided 
that the following integrability condition is satisfied:
\begin{equation}
\label{integrability}
\frac{1}{a_\CD^6}\partial_t \left(\,{\CQ}_\CD \,a_\CD^6 \,\right) 
\;+\; \frac{1}{a_\CD^{2}} \;\partial_t \left(\,\average{\CR}a_\CD^2 \,
\right)\,=0\;.
\end{equation}
Equation (\ref{integrability}) explicitly shows the generic {\it curvature--fluctuation coupling}, i.e. the dynamical 
coupling between the averaged 3--Ricci curvature and the kinematical backreaction term, and it is a genuinely relativistic 
effect (see \cite{buchert:review} for details). Indeed it is in such a fully relativistic, non--perturbative and background--free context where backreaction models have to be studied, if one aims to access their complete properties.

\subsection{Effective cosmological parameters}

In addition to the {\it volume scale factor} $a_\CD$ we introduce the {\it volume Hubble functional}
$H_\CD : = {\dot a}_\CD / a_\CD$, together with the following adimensional average characteristics:
\begin{eqnarray}
\label{omega}
\Omega_{m}^{\CD} : = \frac{8\pi G}{3 H_{\CD}^2} \langle\varrho\rangle_{\cal D}  \;\;;\;\;
\Omega_{\Lambda}^{\CD} :=  \frac{\Lambda}{3 H_{\CD}^2 }\;\;;\;\;\\
\Omega_{\CR}^{\CD} :=  - \frac{\average{\CR}}{6 H_{\CD}^2 }\;\;;\;\;
\Omega_{\CQ}^{\CD} :=  - \frac{{\CQ}_{\CD}}{6 H_{\CD}^2 } \;\;.
\end{eqnarray}
Thus, Eq.~(\ref{averagedhamilton}) assumes the `cosmic quartet relation':
\begin{equation}
\label{hamOmegas}
\Omega_{m}^{\CD}+\Omega_{\Lambda}^{\CD}+\Omega_{\CR}^{\CD}+\Omega_{\CQ}^{\CD}\;=\;1\;\;.
\end{equation}
In the following we shall also refer to the components that are not present in the Friedmannian context as:
\begin{equation}
\label{parameterrelation}
\Omega_X^\CD:=\Omega_{\CR}^{\CD} + \Omega_{\CQ}^{\CD}\;\;.
\end{equation}
If $\Omega_{\Lambda}^{\CD}=0$, then $\Omega_X^\CD$ comprises the Dark Energy contribution, usually dubbed $X-$matter; here, $X-$matter
is decomposed into its physical subcomponents.

Finally, we introduce an effective {\it volume deceleration parameter}:
\begin{equation}
\label{deceleration}
q^\CD \equiv  -\frac{{\ddot a}_\CD}{a_\CD}\frac{1}{H_\CD^2} = \frac{1}{2}
\Omega_m^{\CD} + 2 \Omega_{\cal Q}^{\CD} - \Omega_{\Lambda}^{\CD}\;.
\end{equation} 

\subsection{Correspondence with scalar fields: the morphon field}

It has been shown in \cite{morphon} that the backreaction effect on the kinematics can be mapped into an effective scalar field component $\Phi_{\CD}$,
dubbed ``morphon'', with self-interaction potential $U_{\CD}(\Phi_{\CD})$, such that the backreaction terms comply with the following correspondence:
\begin{equation}
\label{morphoncorres}
-\frac{1}{8\pi G}{\cal Q}_\CD \;=\; \epsilon {\dot\Phi}^2_\CD - U_\CD\;\;\;;\;\;\;
-\frac{1}{8\pi G}\average{\CR}= 3 U_\CD\;\;,
\end{equation} 
where $\epsilon=\pm 1$ is a free parameter which allows the mapping to standard or phantom--like scalar fields. 
In this scheme, we may notice that the effective potential of the morphon field is directly linked to the averaged scalar curvature of space.
Using this correspondence, Eq.~(\ref{averagedraychaudhuri}), (\ref{averagedhamilton}) and (\ref{integrability}) read:
\begin{eqnarray}
3\frac{{\ddot a}_\CD}{a_\CD}&=& -4\pi G \left(\,\average{\varrho}+\varrho_{\Phi}^{\CD}+3p_{\Phi}^{\CD}\,\right) +\Lambda\\
3\left( \frac{{\dot a}_\CD}{a_\CD}\right)^2&=&8\pi G \left(\,\average{\varrho}+\varrho_{\Phi}^{\CD}\,\right)+\Lambda\\
{\ddot\Phi}_\CD + 3 H_{\cal D}{\dot\Phi}_\CD &=&-\epsilon\frac{\partial}{\partial \Phi_\CD}U(\Phi_\CD)\mbox{ ,}
\end{eqnarray}
where $\varrho_{\Phi}^{\CD}\equiv\epsilon\frac{1}{2}\dot{\Phi}_{\CD}^{2}+U_{\CD}$ and $p_{\Phi}^{\CD}\equiv\epsilon\frac{1}{2}\dot{\Phi}_{\CD}^{2}-U_{\CD}$ are the energy density and pressure of the morphon field, respectively. 
This system shows that, at the kinematical level, i.e. as far as the evolution of the effective scale factor is concerned, 
the backreaction effect and a scalar field component cannot be discriminated. Nevertheless, the
backreaction effects are partly of geometrical origin, and in the following we shall see that such a feature is the key point 
that, through the analysis of observations, can discriminate between backreaction and a standard quintessence component.

\smallskip

Overall, our model can be summarized by the following argument: the proposed model is kinematically equivalent (up to the involved dependence on 
the averaging scale) with the standard model supplemented by a quintessence field. The difference is, besides the obvious physical justification of the quintessence field, dynamical: while the standard model has only 
physical sense as a template for the average distribution (since the Universe is not homogeneous), we have to make sure that the exact integral properties are satisfied as well. Hence the alternative model that we propose here can be considered as an attempt to 'repair' the standard FLRW scenario, since this latter is not compatible with the underlying metrical
properties of the low redshift matter distribution, while the former is. Nevertheless, our model maintains some of the naive assumptions of the FLRW case, notably the fact that that light propagates in a constant curvature space, while in reality it travels, most of the time in 4-Ricci flat regions due to the volume--dominance of voids at moderate redshifts \cite{ellisbuchert}.   

\section{Effective geometry}

\subsection{Motivation for the effective metric: construction of distance measurements}
Our approach is the following:
\begin{itemize}
\item[i)] Space--time is foliated using a 3+1 ADM procedure
  \cite{ADM}. Thus the metric reads:
\begin{equation}
\label{truemetric}
^{(4)}{\bf g}=-dt^{2}+g_{ij}(t,x^{k})dX^{i}\otimes dX^{j}\mbox{ ,}
\end{equation}
where $t$ is the cosmic time, latin indices are in $\{1,2,3\}$ and $g_{ij}(t,X^{k})$ is the metric of the spatial hypersurface at time $t$ (first fundamental form), endowed with the local comoving coordinates $X^{k}$.
\item[ii)] An effective averaged model is then built by the averaging
  procedure described in \cite{buchert:grgdust},
  \cite{buchert:grgfluid} and  \cite{buchert:static}. This results in
  the Eqs.~(\ref{averagedraychaudhuri}) and
  (\ref{averagedhamilton})
  which contains the so--called backreaction terms stemming from the averaged
  fluctuations in the local expansion and shear scalars. These equations admit
  solutions for the effective scale factor  $a_{\CD}(t)$, which describes the kinematics of the
  averaged model, provided that the integrability condition Eq.~(\ref{integrability}) is satisfied.
  We shall close this system by the ansatz of exact scaling solutions \cite{morphon} with the scaling index as a free parameter. A
  particular scaling index reproduces the result of second--order perturbation theory in the synchronous gauge \cite{lischwarz}.
\item[iii)] In addition to accounting for the kinematical properties of backreaction models a definition of distance measurements is still
required for a complete description.  So far, the structure of the lightcone has only been explored by
perturbative calculations, e.g. \cite{kantowski1, ruth:luminosity, vanderveldetal}, but our modification of the standard model focuses
  on the effective cosmological evolution and the determination of
  effective distances within a backreaction model that, without any assumptions on the kinematical properties, couples the evolution of
  fluctuations to that of the averaged scalar curvature. This is the reason why our modification is inspired by recent results that deal with
a rescaling of inhomogeneous three--geometries as a possible solution
to the so--called {\em fitting problem} \cite{ellisstoeger, ellisbuchert}.
Ricci flow renormalization of the 
average characteristics on a bumpy geometry (see \cite{klingon,
  buchertcarfora:PRL} and references therein) 
monotonically decreases intrinsic curvature inhomogeneities and would produce a constant--curvature slice, 
{\em but} only at a given instant in time\footnote{In general, such a flow has singularities, if the 3--Ricci tensor is non--positive, and a constant--curvature model is reached only after
subsequent steps of surgery of the manifold. Employing surgery, Perelman \cite{perelman:entropy, perelman:ricci} has added the final proof of the possibility of smoothing a generic 3--manifold into pieces of constant--curvature manifolds with the help of the Ricci flow.}. This result then provides us with a prescription to construct distance measurements along the lightcone. 
As an effective model for the geometry of our Universe, we choose to replace, at a given scale corresponding to the domain $\CD$, the true 3+1 foliation given by the metric, Eq.~(\ref{truemetric}), with another 3+1 foliation such that at each time $t_{*}$, the spatial hypersurface is modeled by a Riemannian 3-space of spatially constant curvature, which we denote as $\kappa_{\CD}(t_{*})$. Such a curvature term can obviously be a function of time $t_{*}$, since there is no reason why an homogeneization procedure such as the Ricci flow, that is applied separately at any time, would result in the same constant curvature space. In contrast, the coupling between spatial fluctuations in the second fundamental form and the smoothed curvature implies that, if the fluctuations have changed from one instant in time to another, the resulting smoothed curvatures at these two instants may be different. In such a case a template model of space--time consists of a family of three--dimensional hypersurfaces $\Sigma_{t}$ of spatially--constant curvature, with each hypersurface being characterized by two scalars: $a_{\CD}(t)$ for the kinematical properties and $\kappa_{\CD}(t)$ for the curvature. Therefore, the model is specified by $\left(\Sigma_{t},a_{\CD}(t),\kappa_{\CD}(t)\right)_{t\in \mathbb{R}}$.  Since the kinematics of an averaged model is governed, in general, by the volume scale factor $a_\CD$ and not by the Friedmannian
scale factor $a$, introducing the former in the effective description of the lightcone is a justifiable assumption (see \cite{singh2})\footnote{Note
here, that the values of these two scale factors are not expected to
differ strongly on large scales. On the other hand, their {\em
  derivatives} can differ significantly and this is accounted for by
allowing for an explicit and consistent 
change of the scalar curvature. As we assume that
  the standard FLRW model describes the Early Universe, including the
  recombination epoch, backreaction effects are assumed to be
  suppressed and consequently both scale factors coincide at early
  times.}.

\end{itemize}

We can now compute the geometrical properties of the effective model. Let us choose on each hypersurface the standard spherical coordinates $(r,\theta,\phi)$. 

At a given time $t_{*}$, the distance between two points on a
  hypersurface $\Sigma_{t_{*}}$ separated by a coordinate distance $r$
  is given by:
\begin{equation}
\label{physdist1}
l(t_{*})=a_{\CD}(t_{*})\int_{0}^{r}\frac{dr'}{\sqrt{1-\kappa_{\CD}(t_{*})r'^{2}}}\mbox{ .}
\end{equation}

Consider a second hypersurface $\Sigma_{t_{*}+dt}$ infinitesimally close to $\Sigma_{t_{*}}$. 
We can write the distance between the two points over this new hypersurface as
\begin{equation}
\label{physdist2}
l(t_{*}+dt)=a_{\CD}(t_{*}+dt)\int_{0}^{r}\frac{dr'}{\sqrt{1-\kappa_{\CD}(t_{*}+dt)r'^{2}}}\mbox{ .}
\end{equation}
By Taylor expanding Eq.~(\ref{physdist2}) around $t_{*}$, keeping only the first order terms in $dt$ and using (\ref{physdist1}), one obtains:
\begin{equation}
\label{modifhubble}
\frac{dl}{dt}(t)=H_{\CD}(t)l(t)+a_{\CD}(t)\frac{d\kappa_{\CD}}{dt}(t)\int_{0}^{r}\frac{r'^{2}dr'}{1-\kappa_{\CD}(t)r'^{2}}\mbox{ .}
\end{equation}
The first term on the right--hand--side is the standard Hubble law, and
the second term expresses a deviation from this Hubble law induced by the particular geometrical structure of the effective model.
Let us consider now the motion of photons. Since their velocity is constant and equal to $c$ along their geodesics, one has the relation:
\begin{equation}
\label{photvel}
\frac{dl}{dt}(t)=c\mbox{ .}
\end{equation}
Hence, using Eq.~(\ref{physdist1}), one finds:
\begin{equation}
\label{photvel2}
cdt=dl(t)=a_{\CD}\frac{dr}{\sqrt{1-\kappa_{\CD}(t)r^{2}}}\mbox{ ,}
\end{equation}
or equivalently, the following differential equation for the coordinate distance travelled by a photon:
\begin{equation}
\label{EDOcoorddistphot}
\frac{dr}{dt}=\frac{c}{a_{\CD}(t)}\sqrt{1-\kappa_{\CD}(t)r^{2}}\mbox{ .}
\end{equation}

\subsection{The template metric}

The ordinary differential equations (\ref{modifhubble}) and (\ref{EDOcoorddistphot}) suggest to introduce an effective metric to describe the geometry in the 4-dimensional template averaged model, so that the photons move on null-geodesics of this metric. These equations are exactly the same as those that would be inferred from the following space--time metric:
\begin{equation}
\label{eq:tempmetric1}
{}^4 {\bf g}^\CD = -dt^2 + L_{\now H}^2 \,a_\CD^2 \gamma^\CD_{ij}\,dX^i \otimes dX^j \;\;,
\end{equation}
where $a_{\now\CD}L_{\now H}=1/H_{\now\CD}$ ($c=1$) is the
present size of the horizon introduced so that the coordinate distance is adimensional, $a_\CD (t)$ is an adimensional 
{\em volume scale factor} on a rest mass--preserving compact domain $\CD$, endowed with a domain--dependent {\em effective three--metric} that assumes the form:
\begin{equation}
\label{eq:tempmetric2}
\gamma^\CD_{ij}\,dX^i \otimes dX^j =\left(\frac{dr^2}{1-\kappa_{\CD}(t)r^2}+d\Omega^{2}\right)\mbox{,}
\end{equation} 
where $d\Omega^{2}=r^2(d\theta^{2}+\sin^{2}(\theta)d\phi^2)$ is the solid angle element.
This form has been suggested, using different considerations, by Paranjape and Singh \cite{singh2}.

At any given time, the template three--metric is identical to the metric of the spatial part of a FLRW space--time, but with a scalar curvature that smoothly varies from time to time.

However, for consistency with the general kinematical properties discussed above, $\kappa_{\CD}$ cannot be arbitrary, and it must be related to the true averaged scalar curvature $\average{\CR}$. We propose, in analogy with a
FLRW metric, to choose $\kappa_{\CD}$ such that:
\begin{equation}
\label{eq:defkappa}
\average{\CR}=\frac{\kappa_{\CD}(t)|\naverage{\CR}|a_{\now\CD}^{2}}{a_{\CD}^{2}(t)}
\mbox{ .}
\end{equation}
This choice guarantees that $\kappa_{\CD}$ does not have any dimension, and
suitably inherits the sign of $\average{\CR}$. We shall discuss the
possibility to test this prescription at the end of the paper. The FLRW case can be recovered by formally posing $\kappa_\CD(t_0)|\naverage{\CR}|=k_{\now\CD}/6$, with $\kappa_\CD(t_0):=1$, where $k_{\now\CD}$ is identified with the constant Friedmannian curvature on a chosen large domain
$\now\CD$.

It is worth emphasizing that this template metric is not required to be a dust solution of Einstein's equations \cite{boundsoncurvature}, \cite{rasanenk(t)} (the effective fluid of an averaged dust model also features a geometrical pressure
\cite{buchert:review}). In fact Einstein's field equations are satisfied locally for a general, unspecified space--time metric, whereas the template metric defined by (\ref{eq:tempmetric1}) and (\ref{eq:tempmetric2}) is only a prescription to take into account the non--trivial behavior of the three--curvature in the averaged effective model of the Universe, and to compute quantities along the approximate smoothed lightcone associated with the travel of light in a clumpy Universe (for another earlier attempt see \cite{palle}).
Recently, Kasai \cite{kasaiSN} has investigated the goodness of fit to supernov{\ae} data
of Friedmannian models without cosmological constant, and with {\em different} curvature
parameters. While a {\em single} standard model without cosmological constant cannot
account for the supernov{\ae} data, two such models -- if applied to low-- and high--redshift
data separately -- can reproduce the observations \cite{kasaiSN}. Kasai also provides in his paper formulae from a post--Newtonian
description of inhomogeneities to explain dynamically the link between these two models.
Our (background--free) model differs in this particular aspect, since
we work with general averaged cosmologies, e.g. we do not assume
the scale factor to be that of a FLRW solution. Also several other
works have concentrated on the role of curvature (however, within 
the class of constant--curvature models), e.g. \cite{franca}, \cite{bruceSN}.

We shall show below that supernov{\ae} data by themselves indicate that the assumption
(\ref{eq:defkappa}) is indeed reasonable.
From now on, we consider this effective metric to compute cosmic distances.

\subsection{Computation of observables}

The computation of distances is a very important issue that must
complement the kinematics when comparing the predictions of a given
averaged model with observations.     
In an inhomogeneous Universe, this is a difficult
problem that can be overcome either by considering exact solutions of Einstein's
equations with particular symmetries such as the LTB metric, or by explicitly
computing the photon path through a perturbed FLRW background \cite{ruth:luminosity,vanderveldetal}. 

The above prescription considerably facilitates the computation of effective distances.
Firstly, let us introduce formally an effective redshift $z_{\CD}$. The redshift is defined as:
\begin{equation}
\label{eq:redshiftdef}
1+z_{\CD}:=\frac{(g_{ab}k^{a}u^{b})_{S}}{(g_{ab}k^{a}u^{b})_{O}}\mbox{ ,}
\end{equation}
where the O and the S stand for the evaluation of the quantities, respectively, at the observer and at the source. In this expression, $g_{ab}$ is the effective metric (\ref{eq:tempmetric1}), $u^{a}$ the 4-velocity of the matter content ($u^{a}u_{a}=-1$), supposed to be comoving hereafter, and $k^{a}$ the wave vector of a light ray travelling from the source S towards the observer O ($k^{a}k_{a}=0$). Then, by normalizing this wave vector such that $(k^{a}u_{a})_{O}=-1$ and introducing the scaled vector $\hat{k}^{a}=a_{\CD}^{2}k^{a}$, one obtains the equation:
\begin{equation}
\label{eq:defred2}
1+z_{\CD}=(a_{\CD}^{-1}\hat{k}^{0})_{S}\mbox{ ,}
\end{equation}
with $\hat{k}^{0}$ obeying the null geodesics equation $k^{a}\nabla_{a}k^{b}=0$ which, since in our case we are only interested in $k^{0}$, leads to\footnote{See the note at the end of the paper.}:
\begin{equation}
\label{eq:evolk}
\frac{1}{\hat{k}^{0}}\frac{d\hat{k}^{0}}{da_{\CD}}=-\frac{r^{2}(a_{\CD})}{2(1-\kappa_{\CD}(a_{\CD})r^{2}(a_{\CD}))}\frac{d\kappa_{\CD}(a_{\CD})}{da_{\CD}}\mbox{ .}
\end{equation}

As usual, the dimensionless coordinate distance can be derived from the
equation of radial null geodesics, as was done for Eq.~(\ref{EDOcoorddistphot}):
\begin{equation}
\label{eq:coorddist}
\frac{dr}{da_{\CD}}=-\frac{H_{\now\CD}}{a_{\CD}^{2}H_{\CD}(a_{\CD})}\sqrt{1-\kappa_{\CD}(a_{\CD})r^{2}}\mbox{
  ; }r(0)=0\mbox{ .}
\end{equation}

Solving Eq.~(\ref{eq:coorddist}) provides the coordinate distance
$\bar{r}(a_{\CD})$ that can then be plugged into Eq.~(\ref{eq:evolk}) to find the relation between the redshift and the scale factor. With both $r(a_{\CD})$ and $z(a_\CD)$, one can determine the angular diameter ($d_{A}(z_{\CD})$) and the luminosity ($d_{L}(z_{\CD})$) distances of sources respectively, 
which read: 

\begin{eqnarray}
\label{eq:distances}
d_{A}(z_{\CD})&=&\frac{c}{H_{\now\CD}}a_{\CD}(z_{\CD})\bar{r}(z_{\CD})\;\;,\\
d_{L}(z_{\CD})&=&(1+z_{\CD})^{2}d_{A}(z_{\CD})\mbox{ ,}
\end{eqnarray}
the last relation holding because of the reciprocity relation between distances.
Having computed these functions, it is then possible to compare the backreaction
model predictions with the data. 

The observations of SN standard candles provide direct measurements of the luminosity distance as
function of the redshift, thus they can be used to constrain the effective parameters (previously introduced) without the need for
any further assumptions on the model. On the other hand the analysis
of the measurements of the angular distribution of the CMB temperature fluctuations
requires the solution of the Boltzmann equation together with the
Einstein equations. So far this has been derived only in the context
of the linear perturbation theory about the FLRW solution. It is
  beyond the scope of this paper to perform this calculation in the
  backreaction context. Nevertheless, as mentioned before, it is
  reasonable to assume that the Early Universe (until the recombination
  epoch) is well--described by a perturbed FLRW model\footnote{In fact we have no other choice until effective
properties of an inhomogeneous inflationary model are analyzed in detail, which would help to decide whether we can identify a
homogeneous--isotropic state at the exit epoch with a homogeneous--isotropic solution, or whether we have to deal with 
a backreaction remnant also in the initial data. We have to admit that our arguments of a coupling of fluctuations with geometrical properties would also 
apply in the inflationary context.}. Therefore, with this assumption, we can
  safely limit our analysis to only using the cosmic distance
  information encoded in the location of the acoustic oscillations in
  the CMB power spectrum (see \cite{Hupeaks, Doranpeaks}, and for a recent
  detailed analysis \cite{Pierstepeaks}). 
 In fact the multipoles of the CMB peaks and
  dips can be determined from the measured anisotropy power spectrum
  without the need of making any assumption on the underlying
  cosmology of the late--time Universe, provided this cosmology is well approximated by a FLRW Universe in the context of General Relativity {\it before and up to the time of recombination}. These multipoles can then
  be compared with the model prediction given by
\begin{equation}
\label{CMB_multipoles}
l_{m}=l_{a}(m-\phi_{m})\mbox{ ,}
\end{equation} 
where $m\in \mathbb{N}^{*}$ for peaks and $m\in\{3/2,5/2,...\}$ for
troughs, and 
\begin{equation}
\label{soundhorizon_angle}
l_{a}\equiv \pi\frac{\bar{r}(a_{\CD,{\rm rec}})}{r_{\rm s}(a_{\CD,{\rm rec}})}\mbox{ ,}
\end{equation}
where $a_{\CD,{\rm rec}}$ is the scale factor at the recombination, $\bar{r}(a_{\CD ,{\rm rec}})$ is the comoving distance between recombination and an
observer located on Earth today (obtained by integrating Eq.~(\ref{eq:distances})), and $r_{\rm s}(a_{\CD,{\rm rec}})$
the sound horizon size at recombination given by multiplying the sound
speed $c_{s}/c$ by the result of the integration of
Eq.~(\ref{eq:distances})) between $a_{\CD,{\rm rec}}$ 
and $a_{\CD}=0$.
The term $\phi_{m}$ parameterizes the effect of dragging
gravitational forces occurring before recombination and which displace
the maxima and minima of the acoustic oscillations with respect to the
sound horizon scale \cite{Doranpeaks}. The amplitude of these shifts only
depends on the matter and baryon densities, on the redshift of recombination, 
on the scalar spectral index for primordial fluctuations and on the
density of Dark Energy at the time of last
scattering. Parameterization formulae for $\phi_m$ are given in \cite{Doranpeaks,Durrerpeaks}. They depend on the following set of parameters: the matter density $\Omega_{m}^{\now\CD}$, the baryon density\footnote{This quantity is defined exactly as $\Omega_{m}^{\now\CD}$ in (\ref{omega}), and $h:=H_{\now\CD}/100$ as usual.} $\Omega_{b}^{\now\CD}h^{2}$, the spectral index $n_{s}$, and the Dark Energy density at the last scattering surface.
Since we have supposed that the Universe is almost Friedmannian at the
time of last scattering and since in our model the density of Dark
Energy is directly related to the deviation from a Friedmannian
universe model, we shall neglect the dependence on Early Dark Energy. 
We also impose that the spectral index $n_s$ is $1$, 
corresponding to a scale invariant primordial power spectrum. 
Then, the evaluation of $\phi_{m}$ only depends on pre--recombination
quantities that we include in our likelihood analysis. We wish to
remark that the only quantity that is specific to our model is
$l_{a}$, which is explicitly computed for the template metric 
(\ref{eq:tempmetric1}) by integrating Eq.~(\ref{eq:coorddist}).
$l_a$ is also used in the alternative approach by
\cite{mukher1}, but their second quantity, 
$R$, cannot be used safely since it depends on the behaviour
of perturbations \cite{mukher2}.

\section{Constraints from supernovae and CMB observations}

We perform a likelihood analysis to infer constraints on the
backreaction models using the SN Ia data from the SNLS collaboration
\cite{SNLS} and the position of the CMB peaks and dips from WMAP3--yr
data measured  in \cite{Hinshaw}. In particular we focus our analysis
on a simple class of backreaction solutions, the so called `scaling solutions', that 
we are going to describe below.

\subsection{Exact scaling solutions} 

Scaling solutions have been extensively studied in \cite{morphon} with $\Lambda=0$. Following the notations of \cite{morphon}, 
we look for solutions to the backreaction problem in the form:
\begin{equation}
\label{eq:scalingsol}
{\cal Q}_{\CD}={\cal Q}_{\initial\CD}a_{\CD}^{p} \mbox{ ; }
\average{\CR}=\laverage{\CR}a_{\CD}^{n}\mbox{ ,}
\end{equation}
where $n$ and $p$ are real numbers. In \cite{morphon}, two types of
solutions were found. The first type, with $n=-2$ and $p=-6$, is not very
interesting for our purpose, since at late time it corresponds
to a quasi--Friedmannian model in which backreaction is completely
negligible. Kinematical backreaction and averaged scalar curvature decouple in this degenerate situation.
This situation corresponds to a `quasi--Newtonian' evolution of fluctuations on a non--dynamical background geometry.
In contrast, the second type of solutions corresponds to a direct coupling
between kinematical backreaction and averaged scalar curvature, for which we find $n=p$, and
${\cal Q}_{\CD}=r\average{\CR}$ with $r=-(n+2)/(n+6)$. Henceforth, we shall consider
backreaction and averaged scalar curvature of the following form, just retaining the generic class of scaling solutions:
\begin{eqnarray}
\label{eq:scal}
\average{\CR}&=&\laverage{\CR}a_{\CD}^{n}\\
{\cal Q}_{\CD}&=&-\frac{n+2}{n+6}\laverage{\CR}a_{\CD}^{n}\mbox{ .}
\end{eqnarray}
Defining
\begin{equation}
\label{normdensparam}
\Omega_{X}^{\CD}=-\frac{2\laverage{\CR}a_{\CD}^{n}}{3(n+6)H_{\CD}^{2}}\mbox{ ,}
\end{equation}
one has:
\begin{eqnarray}
\label{eq:def}
H_{\CD}^{2}(a_{\CD})&=&H_{\now\CD}^{2}\left(\Omega_{m}^{\now\CD}a_{\CD}^{-3}+\Omega_{X}^{\now\CD}a_{\CD}^{n}\right)\\
\kappa_{\CD}(a_{\CD})&=&-\frac{(n+6)\Omega_{X}^{\now\CD}a_{\CD}^{(n+2)}}{|(n+6)\Omega_{X}^{\now\CD}|}\mbox{
  .}
\end{eqnarray}
Then, Eq~(\ref{eq:coorddist}) becomes:
\begin{equation}
\label{eq:coorddist2}
\frac{dr}{da_{\CD}}=\sqrt{\frac{1-\kappa_{\CD}(a_{\CD})r^2}{\Omega_{m}^{\now\CD}a_{\CD}^{-3}+\Omega_{X}^{\now\CD}a_{\CD}^{n}}}\mbox{
    ; }r(0)=0\mbox{ .}
\end{equation}

The leading order of the perturbative estimation of the backreaction effect (i.e the leading order at late time) is included in this class of solutions, 
and corresponds to the case $n=-1$ \cite{lischwarz}. This can be considered as the minimal estimation of the full backreaction effect, since it accounts for
the backreaction in a quasi--Friedmannian Universe, thus very close to a locally homogeneous and isotropic configuration.
For simplicity we assume the full backreaction effect to be described by a power law solution, although
a realistic treatment of the backreaction effect will not correspond to a fixed scaling, since we expect a qualitative change at the epoch of nonlinear structure formation.

In the Dark Energy context, the kinematics described by Eq.~(\ref{eq:def}) corresponds to
introducing a component with constant equation of state
\begin{equation}\label{weqstate}
w^{\CD}=-\frac{1}{3}(n+3)\mbox{.}
\end{equation}
However, we must not forget that the geometry is different due to the time--dependence of
the curvature. The leading order perturbative solution corresponds to $w=-2/3$, while
conversely a cosmological constant (that is a particular morphon and not explicitly introduced) 
would require $n=0$. From the scaling solutions previously described,
we can reconstruct the morphon field; its potential during the matter dominated era is given by \cite{morphon}:
\begin{eqnarray}
\label{morphscalings}
\fl
U(\Phi_\CD)=\frac{-(1+r){\cal R}_{\initial\CD}}{24\pi G}\left((1+r)\gamma_{\CR
  m}^{\initial\CD}\right)^{2\frac{(1+3r)}{(1-3r)}} \,\sinh^{-4\frac{(1+3r)}{(1-3r)}}
\left(\frac{(1-3r)\sqrt{\pi G}}{\sqrt{\epsilon(1+3r)(1+r)}}\Phi_\CD\right)\nonumber\\
\fl
\qquad\quad\,=\, \frac{2(1+r)}{3}\left((1+r)\gamma_{\CR m}^{\initial\CD} \right)^{\frac{3}{(n+3)}} 
\;\langle\varrho\rangle_{\initial\CD}\,
\sinh^{\frac{2n}{(n+3)}}\left(\frac{(n+3)}{\sqrt{-\epsilon n}}\sqrt{2\pi G} \Phi_\CD \right)\;\;,
\end{eqnarray}
where $\langle\varrho\rangle_{\initial\CD}$ is the averaged restmass density of 
dust matter at a given initial time, and $\gamma_{\CR m}^{\initial\CD} \equiv \Omega_{\cal R}^{\initial\CD}/\Omega_m^{\initial\CD}$ is the ratio of energy density in averaged curvature and dust at the same initial time. This potential is well--known in the quintessence context as it corresponds to quintessence fields with a constant equation of state, or in other words, to a constant fraction of kinetic and potential energies \cite{sahnistarobinskii,sahnietal,urenamatos}.

\begin{figure}[h]
\begin{center}
\includegraphics[width=12cm]{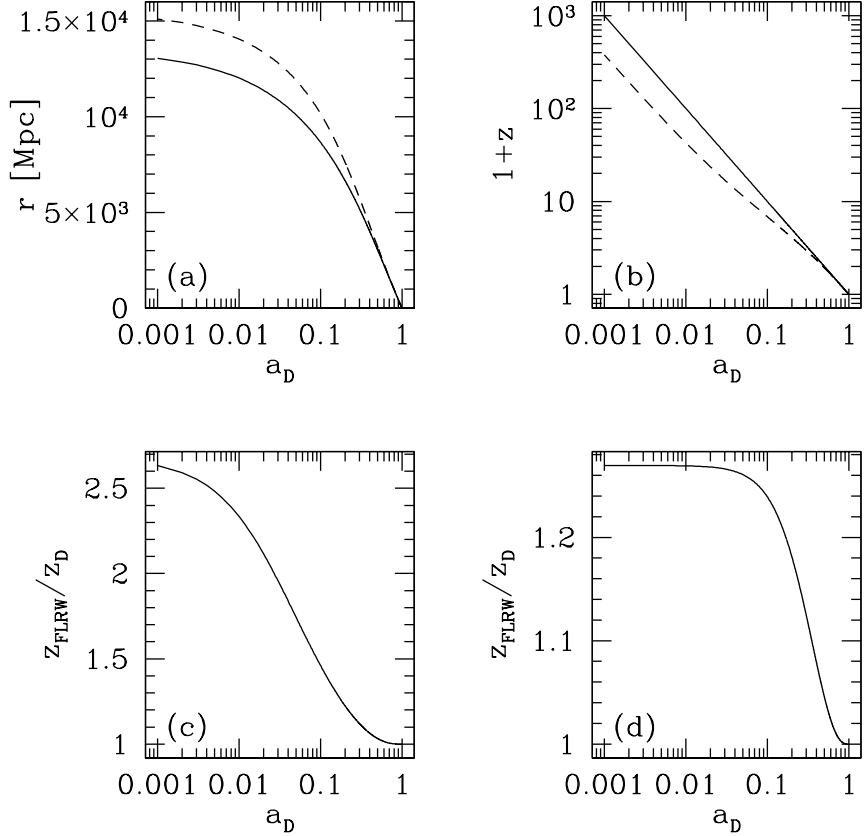}
\caption{\label{fig:rcompare} Evolutions of the coordinate distance $cr/H_{\now\CD}$ (panel a), the redshift (panel b) and the ratio between the FLRW redshift to the redshift, $1/(a_{\CD}(1+z_{\CD}))$ (panel c), in the averaged model as functions of the effective scale factor, in an averaged effective model with $n=-1$, $\Omega^{\now\CD}_{m}=0.3$ and  $H_{\now\CD}=70 \mbox{ km/s/Mpc}$ (dashed line). In Panels a and b, the solid lines represent a FLRW model with the same set of parameters. Panel d represents $1/(a_{\CD}(1+z_{\CD}))$ for the best-fit averaged model described in subsection 4.2.}
\end{center}
\end{figure}

This equivalence shows that, at the level of the dynamical equations for the scale factor $a_{\CD}$, the effect of backreaction cannot be distinguished from the effect of a minimally coupled scalar field in a FLRW context. Nevertheless, the geometrical effects induced by backreaction along the lightcone, and introduced in Subsections 3.1 and 3.2, are a particular feature of a backreaction model and can be a way to distinguish between a FLRW model and a backreaction model. To illustrate this fact, Fig. \ref{fig:rcompare} shows the evolution of the coordinate distance $cr/H_{\now\CD}$ in two models with the same parameters $n$ and $\Omega_{m}^{\now\CD}$, one corresponding to a FLRW Universe, and the other one to a backreaction model with a compatible geometry of the lightcone described by Eq.~(\ref{eq:tempmetric2}) and Eq.~(\ref{eq:defkappa}). It is clear that distance
measures like standard rulers or standard candles can distinguish between the two
models.

\subsection{Constraints}
\begin{figure}[htbp]
\begin{center}
\includegraphics[width=10cm]{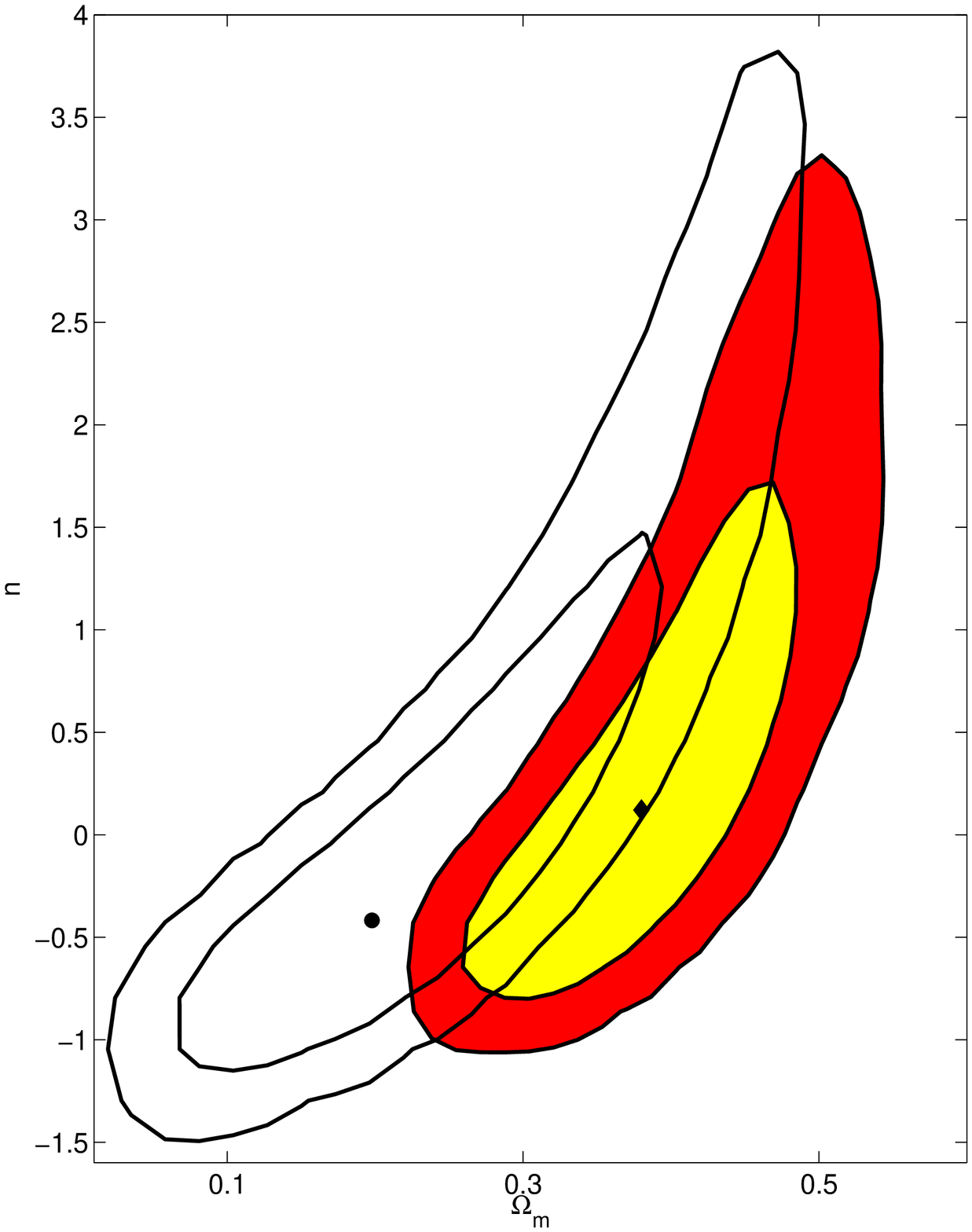}
\caption{\label{fig:ellipse1}Supernovae and CMB constraints in the $(\Omega^{\now\CD}_{m}$,n) plane for the averaged effective model with zero Friedmannian curvature (filled ellipses) and for a standard flat FLRW model with a quintessence
field with constant equation of state $w=-(n+3)/3$ (black ellipses). The disk and diamond represent the absolute best--fit models respectively for the standard FLRW model and the averaged effective model.}
\end{center}
\end{figure}

In Fig. \ref{fig:ellipse1} we plot the joint constraints with $1\sigma$ and $2\sigma$ likelihood contours obtained from supernovae and CMB peak locations for this class of models (filled contours); we also plot the case of a spatially flat FLRW model in presence of quintessence with equation of state given by Eq.~(\ref{weqstate}) (solid lines).

\begin{figure}[h]
\begin{center}
\includegraphics[width=10cm]{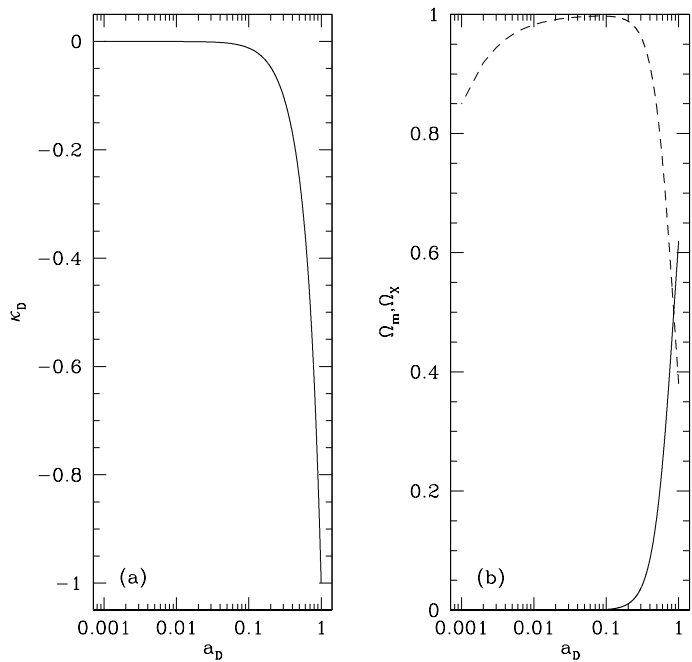}
\caption{\label{fig:param}Left panel: evolution of $\kappa_{\CD}$ for the best--fit averaged effective model (diamond in Fig.~\ref{fig:ellipse1}) with $n=0.12$, $\Omega^{\now\CD}_{m}=0.38$, and $H_{0}=78.54 \mbox{ km/s/Mpc}$. Right panel: evolution of $\Omega_{m}^{\CD}$ (dashed line) and $\Omega_{X}^{\CD}$(solid line) for the same model as in the left panel.}
\end{center}
\end{figure}

It can be seen that the averaged geometry slightly pushes the constraints towards higher values for $\Omega_{m}^{\now\CD}$ compared to standard
Dark Energy models with a FLRW geometry. On the other hand, although a high level of backreaction is needed and acceleration of the effective scale factor still occurs, a lower amount of backreaction is required when accounting for the effective geometry. This is particularly evident regarding the absolute best--fits. The leading perturbative mode ($n=-1$) as calculated in \cite{lischwarz} and \cite{lischwarz2}, is marginally at $1\sigma$ for values of $\Omega_{m}^{\now\CD}\sim 0.3$, whereas a Dark Energy model in FLRW with $n=-1$ is compatible with the data at $1\sigma$ for 
$\Omega_{m}^{\now\CD}\sim 0.1$. However, as expected, the geometrical effect is not sufficient to reconcile the observations with a purely 
perturbative estimate of the backreaction term on horizon scales \cite{lischwarz2}; nevertheless, as a first approximation of a compatible structure of the 
lightcone, the effect acts in the right direction.
In  Fig.~\ref{fig:ellipse1}, we have represented the absolute best--fit for each model: the diamond is for the averaged effective model, corresponding to $\Omega_{m}^{\now\CD}=0.38$, $n=0.12$, $H_{\now\CD}=78.54 \mbox{ km/s/Mpc}$ and $\Omega_{b}^{\now\CD}h^{2}=0.0255$, and the disk is for the flat FLRW model, with $\Omega_{m}^{\now\CD}=0.18$, $n=-0.5$, $H_{\now\CD}=82.0\mbox{ km/s/Mpc}$ and $\Omega_{b}^{\now\CD}h^{2}=0.0263$. The difference in the $\chi^{2}$ between the two models is $3.9$.
The best-fits marginalized over all other parameters (i.e. $H_{\now\CD}$ and $\Omega_{b}^{\now\CD}h^{2}$) are given by $\Omega^{\now\CD}_{m}=0.397$ and $n=0.5$ for the averaged model and by $\Omega^{\now\CD}_{m}=0.26$ and $n=0.24$ for the flat FLRW model. The large difference between the absolute and the marginalized best-fits for the averaged model parameters is caused by the non--Gaussian structure of the likelihood function.
Fig.~\ref{fig:param} shows the evolution of the adimensional density parameters (right panel) and $\kappa_{\CD}$ (left panel) respectively 
for the best--fit backreaction model, corresponding to the diamond in Fig.~\ref{fig:ellipse1} for which $n=0.12$ and $\Omega^{\now\CD}_{m}=0.38$. The ratio between the effective redshift in this best--fit and the standard FLRW redshift, $1/(a_{\CD}(1+z_{\CD}))$, can be found on the panel d of Fig.~\ref{fig:rcompare}. The difference in the early epoch is of order 25 \%.
From the likelihood plot (Fig.~\ref{fig:ellipse1}), we can see that the likelihoods for the two models are displaced along the direction orthogonal to the degeneracy line. For fixed values of $n$ and $\Omega_{m}^{\CD}$ the coordinate distance of the backreaction model is systematically smaller than for the standard Dark Energy model. Hence, in order to fit the data, standard Dark Energy models require more negative values of the equation of state (i.e. larger values of $n$) and smaller values of $\Omega_{m}^{\CD}$ compared to the backreaction models.

\subsection{Testing the curvature prescription (\ref{eq:defkappa})}
In the process of constructing the averaged effective model, we assumed Eq.~(\ref{eq:defkappa}), namely 
that the geometrical instantaneous spatially--constant curvature $\kappa_{\CD}(t)$ is related to the actual averaged scalar curvature $\average{\CR}$. As previously emphasized, this prescription is reasonable from the physical perspective, nonetheless it is worth to test is validity directly
against the data. Indeed, the presence of this non--trivial curvature effect, that makes the effective template metric compatible with the averaged scalar curvature $\average{\CR}$, is purely the result of inhomogeneities. Whereas from a kinematical point of view, acceleration driven by standard quintessence or by backreaction are indistinguishable, a non--trivial curvature effect favoured by the data would be an unambiguous signature of the presence of backreaction. Supposing that the scalings are:
\begin{equation}
\label{diffscal}
\kappa_{\CD}(a_{\CD})\propto a_{\CD}^{m+2}\mbox{ and }\average{\CR}\propto a_{\CD}^{n+2}\mbox{ ,}
\end{equation}
where $n$ and $m$ are arbitrary real numbers, we shall say that the assumption (\ref{eq:defkappa}) is valid if and only if $n \cong m$.
Remember, that a pure scaling ansatz is not what we expect in a realistic evolution of backreaction, so we would be surprised if the data favour this
relation exactly.

\begin{figure}[htbp]
\begin{center}
\includegraphics[width=10cm]{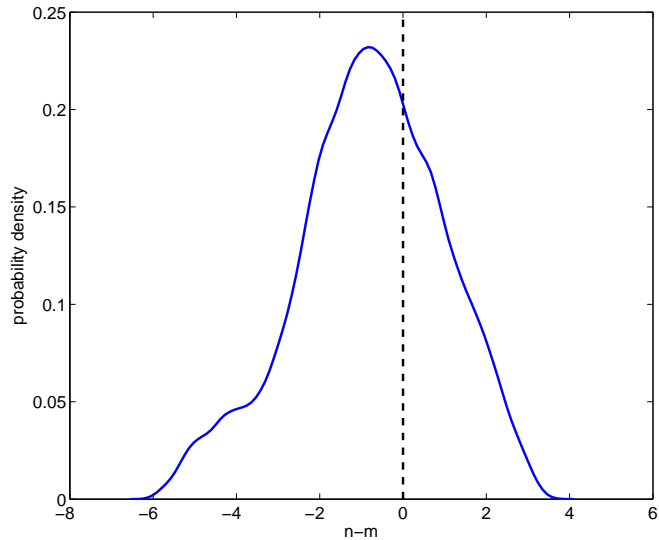}
\caption{\label{fig:pdf} Probability distribution function for the difference of scalings $n-m$, marginalized over all other parameters. The vertical line indicates the case $n=m$ corresponding to Assumption (\ref{eq:defkappa}).}
\end{center}
\end{figure}

Figure \ref{fig:pdf} shows the marginalized likelihood for the variable $n-m$, the vertical line represents the case $n=m$. As we can see, assumption (\ref{eq:defkappa}) lies within $1\sigma$ of the best--fit; it is consistent with current data, but the constraints
are not very strong and currently depend on prior limits for $m$ and $n$ (here we choose both of them to lie in between -4 and 4). It will be interesting to perform this test when more accurate data become available, as well as accounting for better closure conditions of the averaged Einstein equations rather than a simple scaling solution, hence describing the time--evolution of backreaction more realistically.

\subsection{Measuring the evolution of curvature directly}

Another promising way to test Assumption (\ref{eq:defkappa}) relies on a quantity introduced recently in \cite{chris} to test the Copernican principle. In a constant curvature FLRW universe, the coordinate distance $D=r/H_{0}$ is given by:
\begin{equation}
H_0 D(z) = \frac{1}{\sqrt{-\Omega_{k,0}}} \sin\left( \sqrt{-\Omega_{k,0}} \int_0^z \frac{H_0}{H(u)} du \right).
\end{equation}
Taking the derivative of $r(z)$ with respect to $z$ (denoted by a prime in the following)
 turns the sine into a cosine, and it
is easy to see that $(H(z)r'(z))^2 -1 = -\Omega_{k,0} (H_0 r(z))^2$ for any redshift $z$. The
quantity
\begin{equation}
\Omega_k(z) \equiv \frac{(H(z)D'(z))^2-1}{(H_0 D(z))^2}
\end{equation}
is therefore constant and equal to $\Omega_{k,0}$ 
in a FLRW universe. Replacing  $r(z)$ by $\bar{r}(z_{\CD})$, 
and considering the solution to equation (\ref{eq:coorddist}) integrated between $0$ and $z_{\CD}$,
we find that $\Omega^\CD_k(z_\CD)$  is not constant, but
equal to $-\kappa_{\CD}$ as can be seen from Eq.~(\ref{eq:coorddist}).

\begin{figure}[htbp]
\begin{center}
\includegraphics[width=10cm]{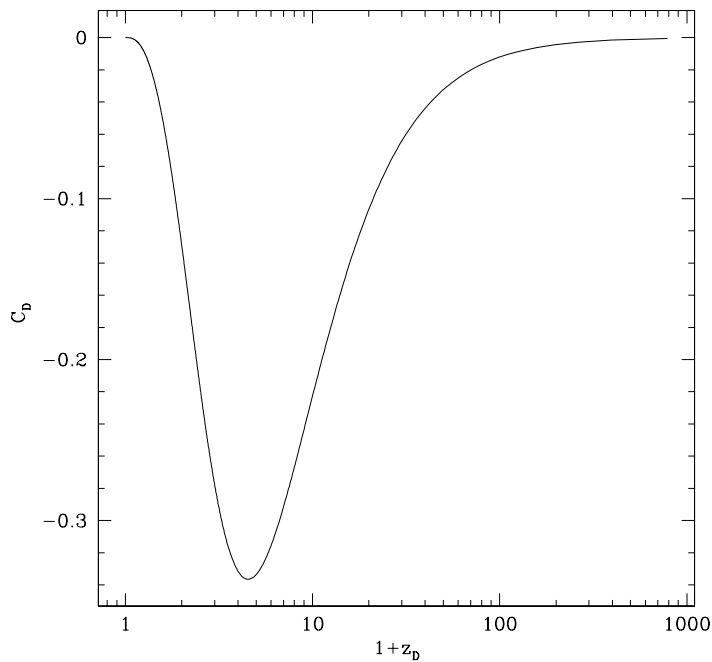}
\caption{\label{fig:coperscal}Evolution of $C_{\CD}$ as a function of $1+z_{\CD}$ for an averaged effective model with scaling backreaction whose parameters are: $n=0.12$, $\Omega^{\now\CD}_{m}=0.38$ and $H_{\now\CD}=78.54 \mbox{ km/s/Mpc}$ (absolute best--fit model represented by a diamond in Fig.~\ref{fig:ellipse1}).}
\end{center}
\end{figure}

Evaluating the derivative of $\Omega_k(z)$, after some algebra we derive the function $C(z)$ defined by equation (3) in \cite{chris}. This function identically vanishes in a FLRW model, whereas for our lightcone prescription it reads:
\begin{equation}
\label{Coperchris}
C_{\CD}(z_{\CD})=-\frac{H_{\CD}(z_{\CD})\bar{r}^{3}(z_{\CD})\kappa_{\CD}^{'}(z_{\CD})}{2H_{\now\CD}\sqrt{1-\kappa_{\CD}(z_{\CD})\bar{r}^{2}(z_{\CD})}}\mbox{ .}
\end{equation}
It is evident that, if the effective geometry of the Universe reduces to FLRW, then $\kappa_{\CD}$ is constant, and $C_\CD$ vanishes for all redshifts, as stated in \cite{chris}. In Fig.~\ref{fig:coperscal} we show $C_\CD$ for the best--fit scaling solution corresponding to the diamond in Fig.~\ref{fig:ellipse1}, that is for $n=0.12$ and $\Omega_{\CD}^{m}=0.38$. There are some features common to all the averaged effective models based on scaling--backreaction that fit the SN1a and CMB data: they all have a $C_\CD$ close to zero at the recombination redshift (corresponding to our assumption that the Universe is well described by FLRW at that time), and a decrease in the late--time Universe, with a minimum at a redshift around $3$ to $4$.

The form of $C_\CD$ or $\kappa_\CD$ in the averaged effective models (\ref{Coperchris}), and its evolution in the best--fit presented in Fig.~\ref{fig:coperscal} can be considered as predictions of the averaged effective model, that could be tested in order to discriminate it from a standard FLRW Dark Energy model. In particular, all models explaining Dark Energy through effects from structure formation will show a strong departure from FLRW at late times, $z_\CD \sim 10$. If this feature does appear in the observational data, it will be an unambiguous sign that the FLRW approximation is not valid in the late--time Universe, and if the behavior of $C_\CD$ follows a curve similar to the one presented in Fig.~\ref{fig:coperscal}, it will indicate that the template metric employed in this paper in conjunction with the backreaction effect are, despite its obvious shortcomings, a satisfactory description of the late--time Universe.

\begin{figure}[htbp]
\begin{center}
\includegraphics[width=10cm,height=15cm]{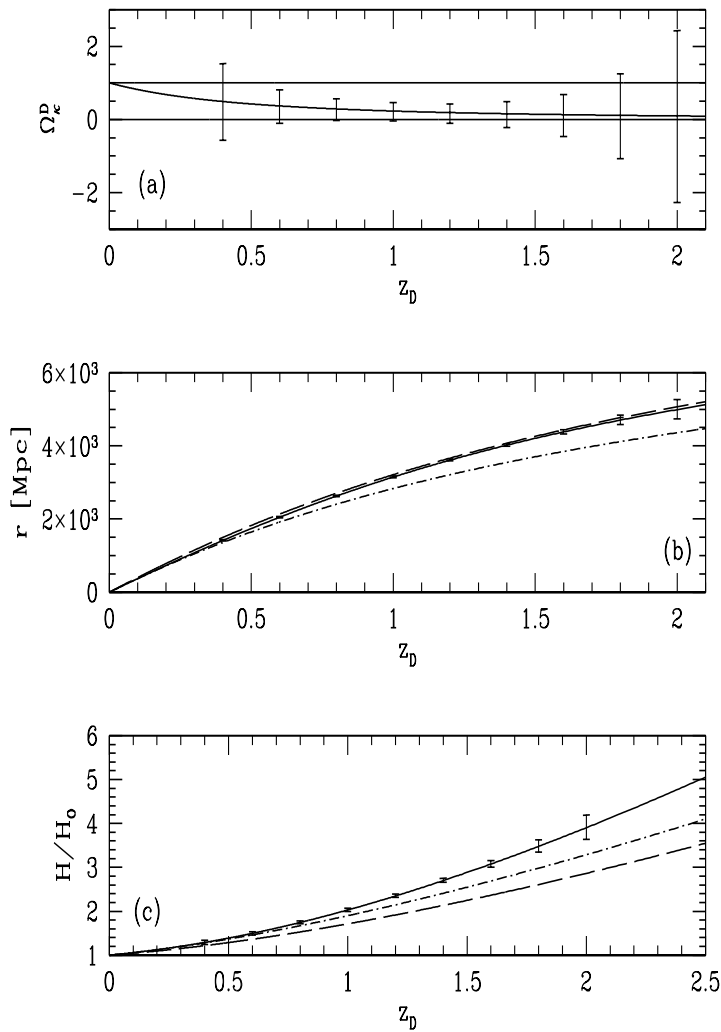}\\
\caption{\label{fig:euclidtest}Upper panel: Evolution of $\Omega_{k}(z_{\CD})$ as a function of redshift for the absolute best--fit averaged model represented by the diamond in Fig.~\ref{fig:ellipse1}. The error bars are obtained using the EUCLID satellite project measure uncertainties. One can see that all positively curved FLRW models ($\Omega_{k,0}<0$) and only highly negatively curved FLRW models ($\Omega_{k,0}>0.5$) can be excluded by the estimation of $\Omega_{k}(z_{\CD})$. Central panel: Evolution of the coordinate distance for the best--fit averaged model (solid line), for a $\Lambda$CDM model with $\Omega_{m,0}=0.277$, $\Omega_{\Lambda}=0.735$ and $H_{0}=73\mbox{ km/s/Mpc}$ (dashed line), and for the FLRW model with the same parameters as the best--fit averaged model (dashed-dotted line). The error bars  are still obtained using the EUCLID expectations. Lower panel: Evolution of the Hubble parameter $H/H_{0}$ for the best--fit averaged model (solid line), the FLRW model with the same parameters as the averaged best--fit model (dashed-dotted line), and for the same $\Lambda$CDM model as in the central panel (dashed line). The error bars are still obtained using the EUCLID expectations.}
\end{center}
\end{figure}

Measuring the function $C_\CD(z_\CD)$ or even $\Omega^\CD_k(z_\CD)$ remains difficult because of the necessity of evaluating the redshift derivative of the coordinate distance directly from the data (see the upper panel of Fig.~\ref{fig:euclidtest}). However, we may notice that, not surprisingly, all necessary information is contained in $H_{\CD}(z_{\CD})$ and $r(z_{\CD})$. In fact, in order to determine the curvature, it is necessary to measure observables which depend on the geometry, such as the luminosity or angular diameter distance, both derived from $r(z_{\CD})$, and another observable that defines the kinematics, i.e. $H_{\CD}(z_{\CD})$. Therefore, to test the geometry, it is sufficient to combine measurements of $r(z_{\CD})$ and $H_{\CD}(z_{\CD})$. If the Universe evolves with a constant curvature, the predictions of the averaged model for $r(z_{\CD})$ and $H_{\CD}(z_{\CD})$ will not fit the data simultaneously; vice versa, if the curvature is not constant, then it will be the FLRW model that badly reproduces the observations. 
Unfortunately there are currently no strong experimental limits on the evolution of the
Hubble parameter. To get an idea of the size of the effect, we will assume that we can use Baryonic Acoustic Oscillations
(BAO) to measure both $H_\CD(z_\CD)$ (from the radial component) and $D_A(z_\CD)=r(z_\CD)/(1+z_\CD)$ (from the transversal component), although there may be some complications (see \cite{chris} for
more discussions). As we shall see shortly, the effect is small and difficult to measure,
so that we consider a future survey extending over a large redshift range, similar to
the one expected for the EUCLID satellite project\footnote{EUCLID is composed of the former proposed DUNE \cite{DUNE} and SPACE \cite{SPACE} missions.}. This survey would be
able to measure the BAO in 9 bins in the range of $z_\CD=0.4$ to $2.0$ with an expected 
accuracy of about $1\%$ in each bin. As shown in Fig.~\ref{fig:euclidtest}, a flat FLRW with a standard quintessence having the same parameters as the averaged cosmology model can be ruled out by EUCLID: even if the two models share the same kinematics, i.e. the same Hubble parameter, they lead to different results for the distance measurement, hence they could be discriminated.

The central and lower panels of Fig.~\ref{fig:euclidtest} show respectively the behavior of the distance and Hubble parameters in the best--fit averaged model (solid line), the FLRW model with the same parameters (dot--dashed line), and a FLRW model with a cosmological constant and the following parameters (dotted line): $\Omega_{m,0}=0.277$, $\Omega_{\Lambda,0}=0.735$ and $H_{0}=73\mbox{ km/s/Mpc}$. The latter model was chosen to reproduce the angular diameter distances of our best--fit averaged model, including the location of the CMB peak. Even if the averaged model correctly describes the Universe, there are FLRW models that can reproduce the distance data, but then cannot account for the Hubble rate, and vice versa. This confirms that EUCLID measurements together with the CMB peak position will be able to discriminate between FLRW models and averaged cosmologies. For a best-fit FLRW model as the $\Lambda$CDM scenario presented here, this will require the use of a combined measurement of the distance-redshift relation and of the Hubble parameter at different redshift, whereas for a FLRW model with the same set of parameters as the best-fit averaged model, a single measurement is sufficient. This latter point is a consequence of the non-trivial relation between redshift and scale factor in the averaged cosmology; indeed, even though the two models share the same Hubble parameter as a function of the scale factor (as a result of their sharing the same parameters), the relation between redshift and scale factor is different in the two models, then resulting in different Hubble rates at a given redshift.

\section{Conclusion and Outlook}

In this paper we have addressed the problem of comparing averaged inhomogeneous cosmologies with observational data by proposing to fit the observations with the help of an improved  template metric, whose form is compatible with the kinematical integral properties of a general averaged model. This template metric has been motivated by the fact that the averaged curvature of space--like hypersurfaces of the space--time foliation is not expected to be constant in time. Indeed, the cosmological principle only requires the {\it spacelike} quantities to be averaged, but does not impose anything on the evolution of these quantities. In other words, the FLRW universe models, which obey a strict cosmological principle, are a very particular subclass of models respecting a weaker cosmological principle presented in this paper. 
We consider the modified template metric as a first approximation tool for interpreting observations in a Universe that appears homogeneous {\it on large scales}, but in which the backreaction effect cannot be neglected. That means that the proper effective lightcone along which the cosmological observations are made cannot be simply approximated by a FLRW lightcone. It is indeed important to notice that this template metric has only been introduced to compute quantities on the lightcone.\\
Thanks to this prescription for the lightcone, we then deduced constraints on the particular class of scaling backreaction, using the supernov{\ae} luminosity/redshift distribution, and the positions of the peaks in the CMB spectrum. We found that the non--trivial geometry of the lightcone induces a slight change in the constraints, with respect to the same models in a FLRW geometry, leading to models compatible with the data for higher values of $\Omega_{m}^{\now\CD}$, that is to say with a smaller amount of backreaction. This is particularly true for the leading perturbative mode ($n=-1$).

One should note that the model presented in this paper still needs an acceleration of the effective volume scale factor to reproduce the data. Recent results based on the the Lema\^\i tre--Tolman--Bondi (LTB) solution
(see \cite{Bolej} and references therein) show that, on the contrary, one can fit the data without an acceleration of the volume scale factor. 
The two results do not necessarily disagree. Indeed, in the LTB model, one fits the data with an inhomogeneous metric having a functional degree of freedom ($t_{b}(r)$ in \cite{Bolej}), and then, one averages the best--fit model to find that there is no acceleration of the effective volume scale factor. 
In this paper, we have first introduced an effective homogeneous model that we fitted to the data, and instead of having a functional dependence in the effective metric, we have made the assumption that backreaction features a scaling behavior with the scale factor. The comparison between the two approaches could be done once more realistic inhomogeneous metrics are at hand, and a more realistic behavior of the backreaction effect could be implemented. Volume acceleration might be a 
consequence of over--evolving the backreaction at early times by a strict scaling ansatz.

One important feature of our results is that the cosmic history and the distances are strongly affected by the introduction of a non--FLRW geometry for the past lightcone, even if the constraints are only slightly different in the $(\Omega_{m}^{\CD},n)$ plane.

Finally, we discussed two ways of testing the assumption (\ref{eq:defkappa}) that links the geometry of the lightcone with the kinematical properties of the model. On the one hand, we found that the supernov{\ae} data are fully compatible with the assumption that the averaged curvature felt by photons along the past null--cone is linked with the averaged Ricci scalar of space--like hypersurfaces according to (\ref{eq:defkappa}). Unfortunately, the current available data are not sufficiently precise to unambiguously show a preferred selection of this hypothesis; this analysis should be done again with future data providing more statistics. On the other hand, we have calculated the explicit form of a function $C_\CD (z_{D})$, previously introduced in \cite{chris} to measure possible violations of the Copernican principle. This form can be considered as a {\it prediction} of the particular models studied in this paper, and it will be a crucial test demonstrating a quantitatively significant difference to the standard FLRW paradigm.
We have shown that the future EUCLID satellite project might be able to distinguish between a FLRW template geometry and the template geometry with evolving curvature in an averaged model  by using joint measurements of a geometrical quantity ($d_{A}(z)$) and a kinematical property ($H(z)$).

The effective metric that we employed has been motivated by physical and mathematical arguments, but it cannot be considered as a fully satisfactory description of the lightcone. In particular, since weak lensing involves a series of local effects of the metric on lightrays rather than just an integrated effect, implementing constraints from weak lensing surveys require a more refined study of the lightcone structure and of its link with spatial averaging. This link can be achieved by implementing the averaging formalism directly on the lightcone; this is the subject of work in progress. Another improvement will come from a closure condition that is better than a simple scaling solution, and that will encode more precisely the time evolution of backreaction. Such a closure condition can be looked for in numerical simulations, analytical approximations (see \cite{marraetal1, marraetal2, rasanenpeaks} for particular approximations, and \cite{wiltshire,wiltshire05,wiltshire07} for an interesting perspective), or observations (see \cite{buchertcarfora:Q} and references therein for remarks on the difficulties of this last approach).  
Moreover, the complete study of other observables like the full CMB spectrum 
is still unavailable and will be crucial for the construction and the test of a `concordance model' for averaged inhomogeneous cosmologies.

\subsection*{\sl Note added during revision}
During the revision of this paper, \cite{Rosenthal} published a preprint in which they pointed out that, in the effective model with a time varying curvature, the redshift can be calculated from first principles, as it is now done in Section 3.3 of this paper.
This refinement of the redshift calculation indeed introduces quantitative changes, actually enhancing the differences between the model presented here and a FLRW model (as compared to the former version of or paper), but does not affect the conclusions of the paper. A few comments are in order to clarify the differences between our analysis and the one proposed in \cite{Rosenthal}. One must note that instead of performing a full MCMC analysis of the model, the authors of \cite{Rosenthal} fitted their model to a $\Lambda$CDM fiducial model. Although it leads to correct models that actually reproduce the data (because $\Lambda$CDM is a very good fitting model), it is by no way guaranteed that such a procedure provides all the acceptable values for the parameters, nor that it provides the most probable ones, as clearly shown by our analysis.
\subsection*{\sl Acknowledgements}
{\small
JL is supported by a Claude Leon Foundation Postdoctoral Fellowship. JL thanks C. Clarkson and B. Bassett for many valuable discussions and for having pointed out the use of $C$ as a test of the Copernican principle. 
TB acknowledges support and hospitality by Observatoire de Paris and Universit\'e Paris 7, as well
as by Universit\'e de Gen\`eve during working visits. 
MK acknowledges funding by the Swiss SNF for part of this work and thanks Syksy R\"as\"anen for valuable discussions.}

\section*{References}

\end{document}